\documentclass[sigconf,nonacm]{acmart}
\usepackage{enumitem}
\usepackage{cleveref}

\newlist{hyp}{enumerate}{2}
\setlist[hyp,1]{before=\itshape,font=\itshape, label=\textbf{Hypothesis} \arabic*:,ref= \arabic*, leftmargin=*}
\setlist[hyp,2]{before=\itshape,font=\itshape, label=\textbf{Hypothesis} \arabic*:,ref= \arabic*, leftmargin=*}

\usepackage[most]{tcolorbox} % ensures enhanced features work
\usepackage{fontawesome5}    % for icons
\usepackage{lmodern}

\usepackage{xcolor}
\usepackage{multirow}

\tcbset{
  designguideline/.style={
    colback=gray!10,
    colframe=gray!60,
    boxrule=0pt,
    leftrule=4pt,
    arc=2pt,
    left=8pt,
    right=12pt,
    top=8pt,
    bottom=8pt,
    enhanced,
    before skip=8pt,
    after skip=8pt,
    width=\linewidth,
  },
  designhypothesis/.style={designguideline},
}

\newcommand{\CalloutBox}[4]{%
  \begin{tcolorbox}[#1]
    \begin{tabular}{@{}p{1.1cm}@{\hspace{4pt}}p{\dimexpr\linewidth-1.1cm-4pt\relax}@{}}
      \multirow{3}{*}{{\fontsize{28}{28}\selectfont #2}}
        & \textbf{#3}\\[2pt]
      & \rule{\linewidth}{0.4pt}\\[2pt]
      & #4
    \end{tabular}
  \end{tcolorbox}}
\newcommand{\GuidelineBox}[2]{\CalloutBox{designguideline}{\faLightbulb[regular]}{#1}{#2}}
\newcommand{\HypothesisBox}[2]{\CalloutBox{designhypothesis}{\faFlask}{#1}{#2}}

\usepackage{booktabs}

\usepackage{amssymb}
\usepackage{caption}
\usepackage{needspace}

\AtBeginDocument{%
  
    \raggedbottom}

\begin{document}

\title[How Decomposing Multi-Modal AI Advice Shapes Appropriate Reliance]{When Modalities (Dis)Agree: How Decomposing Multi-Modal AI Advice Shapes Appropriate Reliance}


\author{Nidhi Mishra}
\affiliation{%
  \institution{Karlsruhe Institute of Technology}
  \city{Karlsruhe}
  \country{Germany}}
\email{nidhi.mishra@kit.edu}
 
\author{Joshua Holstein}
\affiliation{%
  \institution{Karlsruhe Institute of Technology}
  \city{Karlsruhe}
  \country{Germany}}
\email{joshua.holstein@kit.edu}
 
\author{Patrick Bigge}
\affiliation{%
  \institution{Karlsruhe Institute of Technology}
  \city{Karlsruhe}
  \country{Germany}}
\email{patrick.bigge@partner.kit.edu}
 
\author{Gerhard Satzger}
\affiliation{%
  \institution{Karlsruhe Institute of Technology}
  \city{Karlsruhe}
  \country{Germany}}
\email{gerhard.satzger@kit.edu}
 
\author{Philipp Spitzer}
\affiliation{%
  \institution{Karlsruhe Institute of Technology}
  \city{Karlsruhe}
  \country{Germany}}
\email{philipp.spitzer@kit.edu}
 
% If your preamble sets a combined byline for running heads, update it too:
\renewcommand\shortauthors{Mishra et al.}


%\begin{abstract}
%Multi-modal AI systems combine information from different modalities to produce a single piece of advice, yet this fusion obscures how each modality contributes to the final output. Such information could help humans judge when to rely on AI advice, but it remains unclear whether revealing it improves their appropriateness of reliance. We conducted a pre-registered online experiment ($N = 240$) in which participants assessed used-car prices with support from a multi-modal AI system. In a $2 \times 2$ between-subjects design, we varied the availability of modality-level outputs and explanations, while participants encountered both agreeing and conflicting modality outputs. Revealing modality-level outputs increased appropriate reliance, while explanations improved decision accuracy independently. Contrary to our expectations, modality-level advice affected behavior only when modalities agreed, increasing adherence to AI advice rather than helping participants override incorrect advice. These findings suggest that modality-level advice can make AI advice more persuasive without making its errors easier to detect.
%\end{abstract}

\begin{abstract}
Multi-modal AI systems integrate diverse data streams into unified advices, but this fusion conceals how individual modalities contribute to the final output. Consequently, decision-makers cannot observe whether underlying sources aligned or contradicted one another when assessing the provided advice. To test whether revealing this structure supports appropriate reliance, we conducted a pre-registered experiment ($N = 240$) in which participants evaluated used-car prices with multi-modal AI support. In a $2 \times 2$ between-subjects design, we varied the disclosure of modality-level outputs and natural-language explanations, crossed with within-subjects modality agreement. Revealing modality-level outputs increased reliance, while explanations improved decision accuracy independently. However, this occurred only when modalities agreed: visible consensus acted as a persuasive cue that generally amplified adherence, rather than helping participants override errors under conflict. Thus, decomposing multi-modal advice can make AI advice more persuasive without making errors easier to detect.
\end{abstract}



\begin{teaserfigure}
 \includegraphics[width=\textwidth]{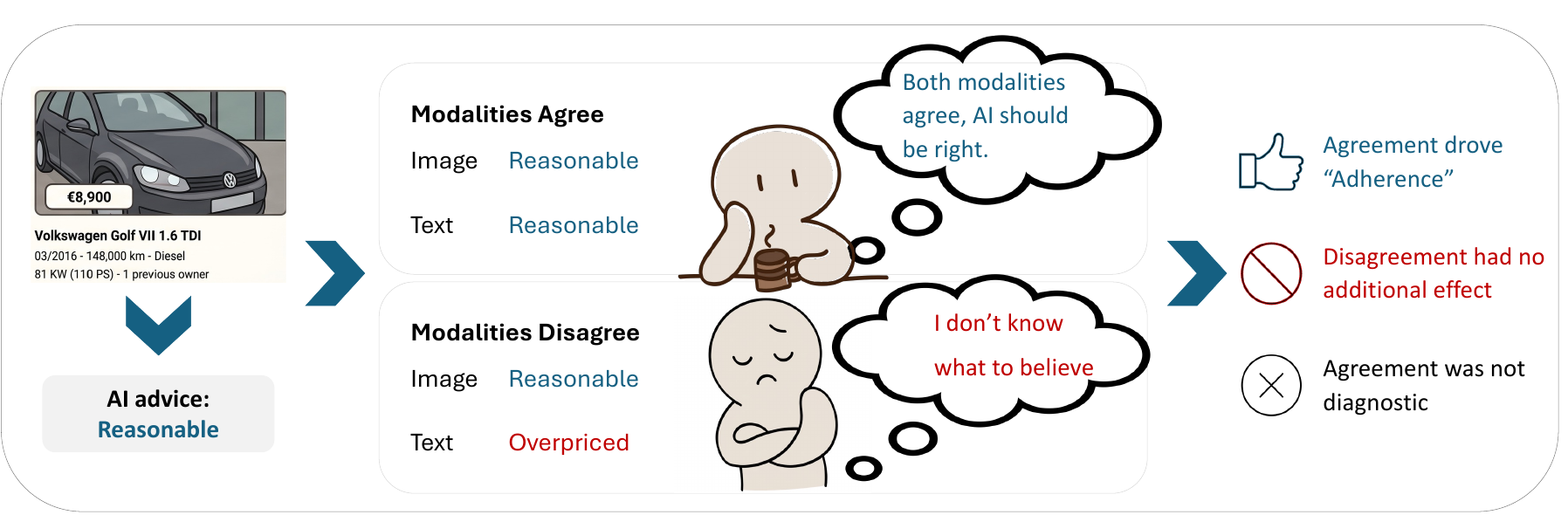}
 \Description{A used-car listing is fused into a single AI advice label of Reasonable. When photo-based and specification-based assessments are also shown, agreement is treated as proof that the fused advice is right, so people follow it more. Visible disagreement leaves decisions unchanged.}
  \caption{When modality-level outputs are revealed, people treat agreement as proof and leave disagreement unused. Participants judged used-car prices with fused AI advice. Showing that the photo- and specification-based assessments agreed increased adherence even though agreement was not diagnostic of correctness ($75\%$ AI accuracy in both strata). Showing that modalities disagreed did not change decisions.}
  \label{fig:teaser}
\end{teaserfigure}



\maketitle

\section{Introduction}

Modern AI systems increasingly operate across multiple input modalities simultaneously, processing, for example, images, text, and structured data within a single model~\cite{liang2024foundations}. In most high-stakes domains, the information relevant to a decision is distributed across data types that no single modality fully captures~\cite{baltrusaitis2018multimodal}. A radiologist's judgment draws on imaging, patient history, and lab results~\cite{yildirim2024multimodal}; a used-car appraisal depends on photographs, textual descriptions, and technical specifications. Multi-modal AI systems combine information from different modalities, allowing them to use complementary information that may improve the accuracy of the final prediction~\cite{liang2024foundations, baltrusaitis2018multimodal}. However, this fusion also hides how each modality assessed the case, leaving humans with only the final prediction.

Effective human-AI collaboration depends on accepting a prediction when it is correct and overriding it when it is wrong, so that the human-AI team outperforms either party alone~\cite{lee2004trust, schemmer2023appropriate, bansal2021does}. Doing so requires information about whether a specific prediction can be trusted~\cite{lee2004trust, zhang2020effect}, but a fused prediction reveals none of the modality-level information that could support the human in doing so. Humans bring their own knowledge and familiarity to different data types, so a buyer assessing a used car may read more from the photographs than from the specifications, or the reverse, depending on what they know about cars. These differences in expertise shape humans' mental models of how the AI should process information and arrive at a advice \citep{holstein2025mentalmodels}. Research in perceptual psychology shows that when humans can distinguish separate information sources, they weight them by perceived reliability in a near-optimal accuracy~\cite{ernst2002humans, fetsch2012neural}. When a multi-modal AI system fuses its inputs, the resulting prediction is typically more accurate, but the modality-level assessments that produced it are no longer visible to the human. Revealing those assessments could therefore provide humans with a basis for deciding when to adhere to the multi-modal AI and when to override it.

Prior work has investigated whether revealing a model's intermediate reasoning to humans improves their decisions \citep{bucinca2021trust, vasconcelos2023explanations}. Showing intermediate concept predictions or visual annotations recalibrated trust in the AI system without reliably improving decision accuracy~\cite{furby2026impact, tschandl2020human}, and the benefits of such are concentrated among humans with less domain expertise~\cite{gaube2023non, tschandl2020human}. These studies, however, operate on a single modality only. These studies, however, operate on a single modality and therefore do not address what happens when information from multiple modalities is fused into one advice. Practitioners working with multi-modal AI diagnostic systems have requested access to this kind of modality-level evidence so they can apply their own expertise to each data source~\cite{yildirim2024multimodal}, but whether providing it changes how appropriately humans rely on the fused advice remains uninvestigated. We therefore ask:
 
\begin{description}
  \item[RQ1.] How does revealing modality-level advice, in addition to overall advice, affect appropriate reliance and human-AI team performance?
\end{description}

Explanations are the most widely investigated mechanism for supporting appropriate reliance in AI-assisted decision-making~\cite{schoeffer2024explanations, schemmer2023appropriate}, though with mixed evidence about their effectiveness~\citep{vasconcelos2023explanations}. Feature-importance explanations have been shown to raise the rate at which participants adopt correct AI advice without lowering the rate at which they override incorrect advice~\cite{schemmer2023appropriate}. In multi-modal AI systems, explanations can articulate why the system reached its prediction, while modality-level outputs show what each information source concluded independently. Both reveal more of the system's reasoning, but whether they complement each other, overlap, or interact remains an open question, which is why we formulate:
 
\begin{description}
  \item[RQ2.] How does the availability of explanations change the effect of revealing modality-level advice on appropriate reliance and team performance?
\end{description}
 
When a multi-modal AI system processes multiple inputs for prediction, its modality-specific predictions do not always reach the same conclusion. For example, in a used-car listing, the image-based and attribute-based assessments may converge or conflict, sending different signals about a car's value. In multi-modal AI, conflicts between modalities are often interpreted as a source of uncertainty and potential unreliability, motivating methods that explicitly detect inter-modality disagreement \citep{christoudias2008multiview,bezirganyan2025multimodal}. Yet the process by which such models adjust their predictions is typically invisible to the human, who receives only the final advice. To understand how (dis)agreement between modalities in multi-modal AI systems affects human decision-makers, we ask: 
 
\begin{description}
  \item[RQ3.] To what extent do these effects depend on whether the multi-modal AI's modalities agree or disagree?
\end{description}

To answer our RQs, we conducted a pre-registered\footnote{We pre-registered our hypotheses, design, and analysis plan on AsPredicted: \url{https://aspredicted.org/ds5b2y.pdf}.} online experiment (N=240) in which participants assessed the prices of used cars with the help of a multi-modal AI system. We used a 2×2 between-subjects design crossing modality-level output disclosure (hidden vs. revealed) with explanation availability (absent vs. present), while systematically varying modality agreement (agreement vs. conflict) within subjects to ensure each participant evaluated instances where the modalities agreed and where they disagreed. We find that presenting modality-level advice increases appropriate reliance, though this falls just short of statistical significance (RQ1). The availability of explanations does not moderate the effect of advice granularity, even though explanations improve appropriateness independently (RQ2). Furthermore, the effect of advice granularity depends on modality agreement. The display alters behavior only under modality agreement and leaves disagreement trials unaffected (RQ3). Decomposing these outcomes shows that modality-specific advice does not improve the ability to distinguish accurate from flawed advice. Instead, it results in indiscriminate adherence, where visible consensus among modalities functions as a persuasive cue. Based on these findings, we make three contributions:

\begin{itemize}
  \item \textbf{Theoretical:} Evidence that modality-level transparency functions primarily as a cue for adherence rather than discrimination. By separating correct adherence from corrective overriding, we show that exposing a multi-modal system's internal assessments increases reliance without improving humans' ability to distinguish correct from incorrect advice, extending recent work of appropriate reliance and transparency in human-AI decision-making.
  \item \textbf{Empirical:} A pre-registered behavioral evaluation, with 240 participants, of revealing the modality-level outputs of a single multi-modal AI system to the human who rely on its fused advice. Revealing the outputs increased adherence to the fused advice regardless of its correctness, and explanations added an independent increment of the same kind.
  \item \textbf{Design:} A stimulus construction that holds multi-modal AI accuracy constant across agreement and disagreement instances, so that reliance shifts can be attributed to the cue itself, together with three design responses that follow from the results, namely revealing modality-level outputs only under conflict, stating what disagreement means for the system's reliability, and explaining the conflict itself in place of each modality's separate case.
\end{itemize}

\section{Background and Related Work}
\label{sec:related} 

Next, we review multi-modal AI systems and the explanation methods developed for them, research on disagreement among modality-specific outputs, and research on appropriate reliance on AI advice.
 
\subsection{Background: Multi-modal AI Systems}

\textbf{Modality contributions are not visible:} Multi-modal AI systems process two or more input channels within a single model \citep{baltrusaitis2018multimodal, liang2024foundations}, on the premise that each carries information the others do not \citep{huang2021makes}. In practice, jointly trained networks often rely on whichever channel is easier to learn and can perform worse than their best single-modality counterpart \citep{huang2022modality, wang2020makes}, most visibly in vision-language models that answer from language priors and largely disregard the image \citep{goyal2017making, agrawal2018dont}. Task accuracy does not reveal which channel carried the prediction \citep{hessel2020does}. Fused advice may therefore rest on one channel without indicating which, and large pre-trained models form their representations before any modality-specific decision exists, so modality-level assessments must be elicited separately rather than read off the model.
 
\textbf{Multi-modal explainable AI:} Two strands of methods recover part of this hidden structure \citep{rodis2024multimodal, joshi2021review}. The first adapts single-modality attribution methods such as SHAP \citep{lundberg2017unified} and LIME \citep{ribeiro2016why} to multi-modal inputs \citep{sotirou2025musiclime}. The second measures each modality's share of a prediction directly, through performance-independent contribution scores \citep{parcalabescu2023mmshap}, joint attributions over regions and tokens \citep{chefer2021generic}, or textual justifications paired with the visual evidence \citep{park2018multimodal}. Attributions computed per modality in isolation can suggest joint reliance where there is little \citep{agarwal2025rethinking}. These methods target model developers and are evaluated against faithfulness criteria, while human-facing evaluation remains rare in XAI \citep{rong2023towards}. Clinicians, for example, accept AI-generated image findings more readily than their synthesis, preferring evidence they can weigh over advice alone \citep{yildirim2024multimodal}. Modality-level outputs can therefore be computed but have not been tested with humans deciding whether to adhere to fused advice.

\begin{table*}[ht!]
\caption{Summary of related work on signals studied in AI-assisted decision-making.}
\label{tab:positioning}
\small
\resizebox{\textwidth}{!}{%
\begin{tabular}{@{}p{0.09\textwidth}p{0.50\textwidth}cccc@{}}
\toprule
Ref. & Summary & MM & Gran. & Disag. & Rel. \\
\midrule
\citep{parcalabescu2023mmshap} & Measures each modality's contribution to a prediction as a model diagnostic. & \checkmark & $\times$ & \checkmark & $\times$ \\
\citep{srikanth2025mixed} & Relates disagreement between uni-modal and fusion models to inter-annotator agreement. & \checkmark & $\circ$ & \checkmark & $\times$ \\
\citep{abtahi2026leveraging} & Proposes preserving minority views across an ensemble of models instead of collapsing them. & \checkmark & $\circ$ & \checkmark & $\times$ \\
\citep{furby2026impact} & Reveals intermediate concept predictions of a concept bottleneck model. Alignment rises without accuracy gains; recalibration requires intervening on the concepts. & $\times$ & \checkmark & $\times$ & \checkmark \\
\citep{tschandl2020human} & Varies granularity of the AI output representation in clinical imaging. & $\circ$ & \checkmark & $\times$ & \checkmark \\
\citep{lu2024second} & Displays a second opinion alongside a primary AI. Disagreement suppresses reliance irrespective of correctness. & $\times$ & $\times$ & \checkmark & \checkmark \\
\citep{tsuchiya2026more} & Displays panels of multiple AI advisors varying in unanimity. & $\times$ & $\times$ & \checkmark & \checkmark \\
\citep{bansal2021does} & Effect of AI explanations on complementary team performance. Explanations raise acceptance of correct and incorrect advice alike. & $\times$ & $\times$ & $\times$ & \checkmark \\
\citep{schoeffer2024explanations} & Decomposes trial-level behavior into adherence and overriding across explanation types. & $\times$ & $\times$ & \checkmark & \checkmark \\
\citep{schemmer2023appropriate} & Conceptualizes appropriate reliance as two-dimensional and tests feature-importance explanations. & $\times$ & $\circ$ & \checkmark & \checkmark \\
\citep{kim2025fostering} & Vary AI response granularity and expose humans to explanations with inconsistencies. & $\times$ & \checkmark & $\circ$ & \checkmark \\
\citep{ma2025towards} & Elicits dimension-level opinions and lets human and system deliberate over conflicts between them. & $\times$ & \checkmark & \checkmark & \checkmark \\
\midrule
 & Our work reveals modality-level advice from one multi-modal AI system, cross-referenced with explanations, with agreement held as non-diagnostic. & \checkmark & \checkmark & \checkmark & \checkmark \\
\bottomrule
\end{tabular}%
}
\vspace{2pt}

{\footnotesize MM = multi-modal system, Gran. = granularity manipulated, Disag. = disagreement between outputs displayed, Rel. = reliance measured with human participants. $\times$ = not addressed, $\circ$ = implicitly addressed or discussed, \checkmark{} = explicitly addressed.}
\end{table*}
 
\subsection{Modality Disagreement}
 
\textbf{Disagreement in multi-modal machine learning:} Divergence between single-modality and fused predictions is treated as a sign of reduced prediction quality. In empathy detection, the examples on which single-modality and fusion models disagree are those on which annotators are uncertain \citep{srikanth2025mixed}, linking model disagreement to instance difficulty \citep{peterson2019human, spitzer2023perception}. Evidential fusion methods build on the same idea, measuring per-modality evidence and treating conflict as an explicit quantity during aggregation \citep{han2021trusted, han2023trusted, xu2024reliable, zhang2023provable}. Throughout, conflict is used within the model to reweight, abstain, or route inputs, and is neither shown to the human nor resolved neutrally \citep{deng2025words}. MEDLEY instead keeps minority views rather than collapsing them into a consensus \citep{abtahi2026leveraging}, but is a conceptual framework and concerns disagreement between separate models, not between modalities within one.
 
\textbf{Human integration of conflicting cues:} Humans have been shown to weigh cues from separate channels in inverse proportion to the reliability of each, for visual and haptic estimates \citep{ernst2002humans} and for audio-visual localization \citep{alais2004ventriloquist}. When cues diverge strongly, humans stop combining them and treat them as separate causes \citep{kording2007causal, shams2010causal}. Yet, this assumes that separate estimates are available, and that humans have experienced each cue's reliability directly.
 
\textbf{Disagreement between AI outputs:} In human-AI interaction, displayed disagreement appears to act as a general trust cue rather than as case-level evidence. A disagreeing second opinion reduces reliance whether or not the AI is correct \citep{lu2024second}, and in multi-advisor panels unanimity produces overreliance while wide disagreement produces hesitation \citep{tsuchiya2026more}, mirroring advice taking, where confidence in an aggregate falls as advisor spread widens \citep{budescu2000confidence, budescu2003effects}. Displayed disagreement does not need to reflect real conflict, since post-hoc explanation methods often produce conflicting signals of the same model and instance \citep{krishna2022disagreement}. One exception suggests the discriminating case looks different, since inconsistencies \emph{within} a single explanation reduced reliance specifically on incorrect outputs \citep{kim2025fostering}. In these studies, disagreement arose between distinct systems, advisors, or runs rather than among modality-level components of one system, and co-occurred with lower advice quality, confounding the two.
 
\subsection{Appropriate Reliance}
 
\textbf{Defining and measuring reliance:} The framing comes from advice taking \citep{bonaccio2006advice, yaniv2004receiving} and trust in automation \citep{lee2004trust}, where the goal is calibration, or for human-AI teams, complementarity, meaning joint performance above what either party reaches alone \citep{hemmer2025complementarity}. That goal is often missed. Across 106 experiments, human-AI combinations performed worse on average than the better of human or AI alone \citep{vaccaro2024combinations}, attributed in part to complacency and automation bias \citep{parasuraman2010complacency}. Team accuracy cannot diagnose such failures, since the same gain follows either from adopting correct advice more often or from discriminating better between correct and incorrect advice \citep{schemmer2022should, schoeffer2025reliance}. Two decompositions separate these, distinguishing adoption of correct advice from retention of one's own correct judgment against incorrect advice \citep{schemmer2023appropriate}, and dividing trial-level behavior into correct adherence, detrimental adherence, corrective overriding, and detrimental overriding \citep{schoeffer2024explanations}.
 
\textbf{What decomposed measures revealed about explanations:} Measured this way, explanations perform less well than their adoption suggests. A meta-analysis found a positive effect of AI assistance on performance and no additional effect of explanations over predictions alone \citep{schemmer2022meta}. Explanations raise adherence without improving discrimination between correct and incorrect advice \citep{bansal2021does, schoeffer2024explanations, kim2025fostering}, can increase overreliance relative to showing the prediction alone \citep{vasconcelos2023explanations}, and mislead when imperfect \citep{spitzer2026dont, morrison2024impact}. One response adds cost to accepting the advice, as in cognitive forcing, which reduces overreliance at the price of higher effort \citep{bucinca2021trust}. Another question is whether explanations are the operative variable, since reliance is better predicted by whether a human can verify a claim independently \citep{fok2024search, chen2023understanding}, and an explanation is better treated as evidence the human weighs than as support for a conclusion the system has already reached \citep{miller2023explainable}.
 
\textbf{Finer-grained advice as an alternative:} One alternative is to give humans the intermediate assessments the advice was built from and let them judge those directly. Routing a prediction through human-interpretable concepts makes such assessments available \citep{koh2020concept}, and revealing them increased alignment with the model without accuracy gains, with alignment tracking model performance only where humans could intervene on the concepts \citep{furby2026impact}. In clinical imaging, more detailed output representations helped less-experienced humans most \citep{tschandl2020human}, and eliciting opinions at the level of individual decision dimensions and discussing conflicts improved appropriate reliance over a conventional XAI baseline \citep{ma2025towards}. Benefits are uneven and depend on expertise and on whether the intermediate level was attended to. A related response reports how likely the advice is to be correct rather than what it was built on, through displays of confidence or uncertainty \citep{pouget2016confidence, zhang2020effect, bhatt2021uncertainty, ma2023who, prabhudesai2023understanding, marusich2024using, holstein2025balancing}. In these studies, the signal reached the human as a summary of overall prediction quality, not localized to parts of the input. Effects across both responses are moderated by domain expertise \citep{gaube2021susceptibility, gaube2023non}, humans' assessment of their own competence \citep{he2023knowing}, and task uncertainty \citep{salimzadeh2024dealing}.

\Cref{tab:positioning} summarizes this work along the dimensions relevant to modality-level transparency. Studies that manipulate granularity do so within a single modality, studies that display disagreement do so between separate systems, and neither line separates the visibility of disagreement from the lower advice quality that usually accompanies it. Our work addresses this gap by revealing the image-based and attribute-based advice of a multi-modal system alongside its fused advice, crossing this with the availability of explanations, and varying modality agreement on stimuli where agreement carries no information about whether the system is correct.

\section{Theoretical Development}
\label{sec:theory}

Research on advice-taking demonstrates that decision-makers evaluate guidance selectively, as the quantity, structure, and origin of advice signals dictate how reliance unfolds and whether final decisions improve \citep{bonaccio2006advice}. When human judgment and AI advice align, initial beliefs require little calibration, but when they conflict, decision-makers must weigh external advice against their own assessments. Multi-modal AI systems introduce unique dynamics to this trade-off because each input stream is processed through dedicated modality components before being fused into a unified advice. In these systems, cross-modality agreement provides a vital, internal proxy for advice reliability, yet this consensus structure is typically hidden from humans behind a single fused advice. We examine how revealing this internal advice structure, alongside explanatory rationales, shapes the appropriateness of reliance across four formal hypotheses.

\textbf{Advice Granularity:} The overall advice collapses agreement or conflict between the modality-specific assessments into a single advice, so the decision-maker sees the system's conclusion but none of the structure that produced it. Appropriate reliance requires cues that vary from case to case, since a decision-maker who receives the same kind of output on every trial has no basis for adhering to the advice on one trial and overriding it on the next~\citep{schemmer2023appropriate, zhang2020effect}. Existing research has shown that finer-grained output can shift AI-supported decisions in uni-modality settings~\citep{tschandl2020human, gaube2023non, furby2026impact}. Decomposing advice into modality-level components provides multiple separate assessments of the same instance grounded in distinct evidence, allowing decision-makers to evaluate and weight each information source by its perceived reliability \citep{ernst2002humans}. When the two assessments point in the same direction, the decision-maker has grounds to follow the advice; when they diverge, they have grounds to critically examine the case before adhering to it. We therefore expect that revealing modality-level advice raises the rate at which decision-makers adhere to correct advice and override incorrect advice.

\begin{hyp}[resume, wide, leftmargin=0cm, labelindent=0pt, labelwidth=0em]
\item Revealing modality-level advice increases appropriate reliance compared with overall advice only.
\label{hyp1}
\end{hyp}

\textbf{Modality Agreement as a Moderator:} The diagnostic value of decomposing fused advice depends on whether the underlying modalities converge, since granular outputs merely confirm an already aligned advice when modalities agree, whereas they reveal an otherwise masked internal conflict when modalities disagree. A fused advice presents itself as an undivided verdict. Making its internal agreement explicit therefore adds little to what the display already implies, while making its internal conflict explicit surfaces information that fusion actively suppresses. Visible inconsistency of this kind functions as an instance-level uncertainty cue that selectively curbs overreliance on flawed advice~\citep{kim2025fostering, marusich2024using}, although conflict displayed between separate systems can also induce indiscriminate skepticism if humans overgeneralize the signal~\citep{lu2024second}. We therefore expect the diagnostic benefit of decomposition to focus on disagreement trials, where decision-makers would otherwise remain unaware of the system's internal disagreement. We therefore hypothesize:

\begin{hyp}[resume, wide, leftmargin=0cm, labelindent=0pt, labelwidth=0em]
\item Modality (dis)agreement moderates the effect of revealing modality-level advice on appropriate reliance, with the effect being stronger on disagreement trials than on agreement trials.
\label{hyp2}
\end{hyp}

\textbf{Explanations:} While granularity provides decision-makers with additional signals to consider, explanations provide them with the analytical foundation needed to determine whether a given signal is trustworthy \citep{bansal2021does, bucinca2021trust}. Without explanations, individuals can observe only what the system recommends, not the rationale driving it, leaving them to adhere to or override the advice based primarily on surface alignment with their prior intuitions. In contrast, providing explanations of the AI's reasoning allows decision-makers to inspect the validity of the underlying logic and calibrate their reliance accordingly~\citep{lai2019predictions, schemmer2023appropriate}. Since case-level rationales provide the structural transparency necessary to detect reasoning breakdowns, we expect their availability to improve reliance calibration. Therefore, we formulate:

\begin{hyp}[resume, wide, leftmargin=0cm, labelindent=0pt, labelwidth=0em]
\item Providing explanations for AI advice increases appropriate reliance compared to not having explanations.
\label{hyp3}
\end{hyp}

\textbf{Explanations and Granularity Together:} The usefulness of decomposing advice into modality-level judgments hinges on whether decision-makers can diagnose why component modalities diverge. Without explanations, a disagreement leaves humans aware of an internal conflict without the means to evaluate it, thereby reducing the granular communication of modality-level advice to an opaque signal. With explanations accompanying each modality-level judgment, decision-makers can inspect the evidence behind the conflict and evaluate the substantive basis of each advice. Since rationales enable humans to scrutinize conflicting advice signals instead of just observing them~\citep{bansal2021does}, explanations should amplify the diagnostic benefits of advice granularity. We therefore hypothesize:

\begin{hyp}[resume, wide, leftmargin=0cm, labelindent=0pt, labelwidth=0em]
\item Providing explanations for AI advice moderates the effect of revealing modality-level advice on appropriate reliance, with the effect being stronger when explanations are present than when they are absent
\label{hyp4}
\end{hyp}

\section{Experiment Design}
\label{sec:experiment_design}
To examine how advice granularity and explanation availability shape appropriate reliance, we conducted an IRB-approved, pre-registered behavioral experiment ($N = 240$) using a used-car price-reasonableness task. We implemented a $2 \times 2$ between-subjects design (advice granularity $\times$ explanation availability) crossed with a within-subjects modality agreement factor. Below, we detail the task environment, AI generation pipeline, experimental conditions, and evaluation metrics.

\begin{figure*}[h!]
    \centering
    \includegraphics[width=\textwidth]{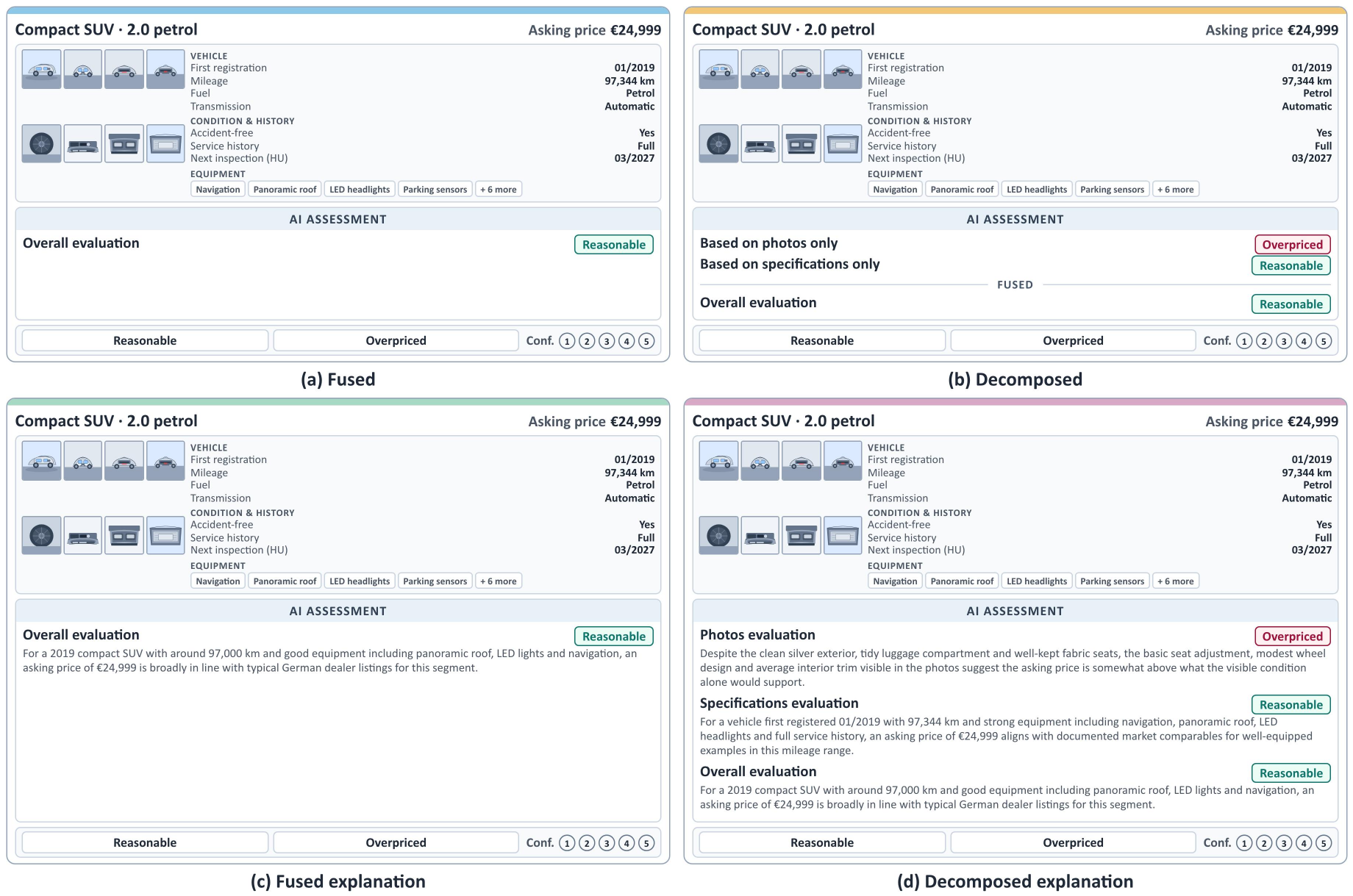}
    \Description{Four-panel mock-up of the participant decision screen. The used-car listing is the same in every panel. Panel (a) shows only the fused Reasonable/Overpriced advice; (b) adds separate photo and specification advice on a disagreement trial; (c) adds a natural-language explanation to the fused advice; (d) pairs each of the three advice lines with its own explanation.}
    \caption{Stylized participant interface under the four between-subjects conditions (advice granularity $\times$ explanation availability). Listing photographs, attributes, and asking price are identical across panels; only the AI assessment block changes. (a)~fused advice; (b)~fused plus per-modality advice, shown here on a disagreement trial; (c)~fused advice with explanation; (d)~per-modality and fused advice, each with an explanation. The study was administered in German; English is used here for readability.}
\label{fig:task_interface}
\end{figure*}

\textbf{Task and Dataset:} Participants judged whether the asking price of a used-car listing was reasonable or overpriced, given the vehicle's photographs, its structured attributes (brand, model, registration year, mileage, fuel type, and equipment features), and the seller's asking price. We chose this task because photographs and attributes each carry information the other lacks, so the two modalities can support or contradict one another on a given listing, and because human with ordinary used-car experience can form their own judgment from both sources without professional training. We scraped listings from mobile.de in June 2026 and retained those with complete photographic and technical records (\Cref{app:dataset_sampling}). Three independent automotive experts established the ground truth for each listing without access to mobile.de's own price rating or to the AI advice. On each trial, participants examined one listing alongside the AI advice for their condition and recorded a binary decision of Reasonable or Overpriced (\Cref{fig:task_interface}).

\textbf{AI System:} We generate all AI advice before data collection using GPT-5.1 at temperature~0, prompting the model to act as an experienced German automotive appraiser (see \Cref{app:ai_pipeline} for the prompt). For each listing, we queried the model through three independent calls to produce genuinely separate modality evaluations instead of post-hoc algorithmic combinations. The image-only call evaluated the asking price alongside up to four evenly sampled photographs; the attribute-only call assessed the asking price alongside structured vehicle specifications while withholding platform ratings and dealer identities; and the combined call integrated photographs, specifications, and asking price into overall advice. Each query generated a binary advice of ``Reasonable'' or ``Overpriced'' alongside a concise natural-language explanation. We restricted these explanations to merely argue for the available modality, i.e., requiring image rationales to draw exclusively on visible vehicle condition while confining attribute rationales to technical specifications.

\textbf{Instance Selection:} To create the experimental stimulus set, we constructed a stratified pool of 64 vehicle listings, balancing across the four combinations of modality agreement (agree versus disagree) and ground truth (reasonable versus overpriced). Within each of the four quadrants, we selected three AI-correct listings for every one AI-incorrect listing, which fixed AI accuracy at exactly 75\% across both the agreement and disagreement strata. This balance ensures that any behavioral shift observed when revealing modality disagreement stems directly from the disagreement cue, not from underlying accuracy disparities between strata. From this pool, each participant evaluated 16 listings consisting of four randomly drawn instances from each quadrant presented in counterbalanced order. We report candidate filtering counts alongside full pool and trial stratification matrices in \Cref{tab:stimulus_pool_distribution,tab:participant_trial_sampling} in the Appendix.

\textbf{Conditions:} We randomly assigned participants to one of four between-subjects conditions resulting from a 2~$\times$~2 factorial crossing of advice granularity and explanation availability. This structure isolates the main effects of granularity and explanation, while the crossed within-subjects agreement factor allows us to examine each effect separately across agreement and disagreement trials. The resulting four experimental conditions are as follows.

\begin{itemize}
    \item \textbf{Fused:} Participants see only the AI's overall advice, with no explanation.
    \item \textbf{Decomposed:} Participants see the overall advice supplemented by separate image-based and attribute-based advice, with no explanation.
    \item \textbf{Fused Explanation:} Participants see the overall advice together with the combined explanation.
    \item \textbf{Decomposed Explanation:} Participants see the image-based, attribute-based, and overall advice, each paired with its own explanation, with the fused explanation identical in wording to the one used in the Fused Advice + Fused Explanation.
\end{itemize}

\begin{figure}[h!]
    \centering
    \includegraphics[width=\linewidth]{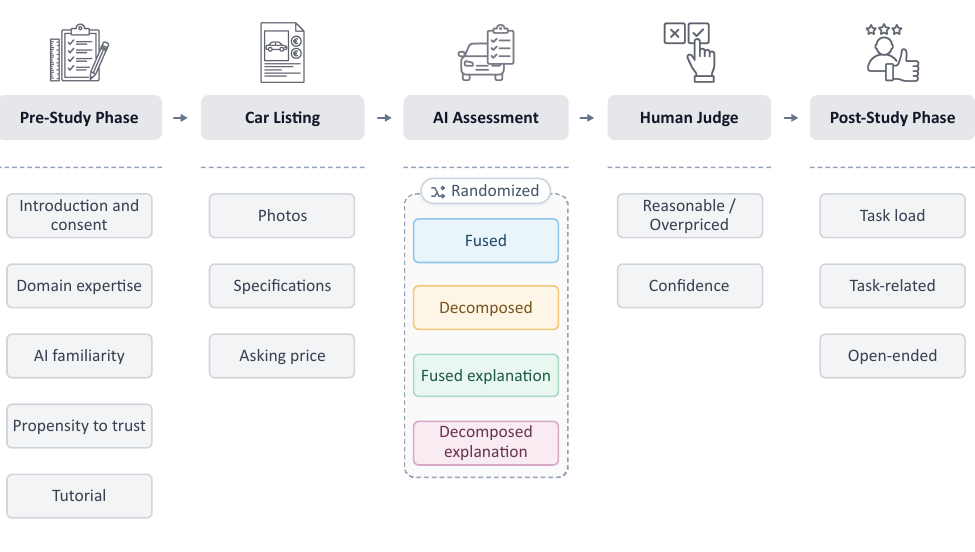}
    \Description{Flow diagram of the experiment procedure, from recruitment and consent through the tutorial and comprehension check, the sixteen decision trials, and the post-task questionnaire.}
    \caption{Procedure of the behavioral experiment.}
    \label{fig:procedure}
\end{figure}

\textbf{Study Procedure:} As illustrated in \Cref{fig:procedure}, the experiment proceeded across three consecutive phases. In the pre-study phase, participants read an introduction, provided informed consent, and completed a questionnaire that captured demographic and control measures. A guided tutorial then introduced the task interface, followed by a comprehension check that allowed up to two attempts, including three questions on the task and AI advice and one on modality-level advice where applicable. We excluded participants who failed the comprehension check twice.  During the main task phase, participants completed 16 decision trials in which they evaluated vehicle photographs and attributes alongside asking prices, together with AI assistance based on their condition. Since the listings came from the German market, we conducted the study in German, with listing attributes, AI advice, and all instructions and questionnaires presented in German. Finally, they decided whether the asking price was ``Reasonable'' or ``Overpriced''. To ensure engagement with the stimuli and prevent premature responses, the interface enforced a minimum viewing time of 20 seconds per listing before allowing participants to move to the next trial. The experiment concluded with a post-study questionnaire in which participants evaluated their subjective experience and answered up to two open-ended questions about their use of the provided stimuli.

\textbf{Metrics:} Our primary trial-level dependent variable is appropriate reliance, operationalized as the correctness of the participant's final decision against ground truth~\citep{schemmer2022should}. In a binary task with fixed AI advice, this indicator is equivalent to what the literature calls human-AI team performance, because agreeing with correct AI advice and overriding incorrect AI advice are the only routes to a correct decision~\citep{schemmer2022should, schoeffer2025reliance}. Following our pre-registered exploratory analysis, we decompose appropriate reliance into four component channels~\citep{schemmer2022should, schoeffer2025reliance}. Correct adherence (adhering to correct AI advice) and corrective overriding (overriding incorrect AI advice) contribute to appropriate reliance, while detrimental adherence (adhering to incorrect AI advice) and detrimental overriding (overriding correct AI advice) work against it. We further distinguish between instances where the AI is correct and those where it is incorrect. In addition to these behavioral metrics, we incorporate pre-task control measures of car-domain expertise, general AI familiarity, and propensity to trust automated systems. In the post-task assessment, we record subjective evaluations of task load \citep{hart1988development}, decision confidence, and trust in the AI system \citep{merritt2013trust}. We report the full psychometric scales, individual survey items, and mathematical formalizations for all measures in \Cref{tab:all_measures} in the Appendix.

\textbf{Participants.}
Following our pre-registration, we recruited participants via Prolific, restricting eligibility to adults residing in Germany or Austria who were fluent in German and reported prior experience purchasing or evaluating used vehicles. We excluded professional car dealers and automotive appraisers to avoid ceiling effects. We excluded 15 participants: 12 for failing the post-tutorial comprehension check on both permitted attempts and 3 for not completing all 16 trials. (\Cref{app:exclusions} reports the full participant flow and the attention-check and honeypot procedures). The final sample comprised 240 participants, 60 per condition  ($M_{\text{age}}=35.4$ years, 65.9\% male, 33.6\% female, 0.5\% non-binary), who completed the study in a median of 17.0 minutes for a fixed payment of \pounds2.50 (\pounds8.82 per hour).

\section{Results}
\label{sec:results}

In this section, we report the empirical findings from our pre-registered experiment by estimating a generalized linear mixed model (GLMM) across all trial-level observations and evaluating our four hypotheses as planned contrasts of that model. For each experimental effect, we decompose appropriate reliance into the two behavioral mechanisms that drive it, specifically, adhering to the AI when it provides correct advice and overriding the AI when it errs. \Cref{subsec:pattern} delineates the overarching descriptive pattern this decomposition reveals across all four conditions. \Cref{subsec:approach} details our statistical model, before \Cref{subsec:granularity}, \Cref{subsec:explanations}, and \Cref{subsec:agreement} analyze advice granularity (H1), natural-language explanations (H3, H4), and modality agreement (H2).

\subsection{The Overall Pattern of Reliance}
\label{subsec:pattern}

Appropriate reliance rises monotonically across the four conditions, increasing from $61.8\%$ under overall advice alone (C1) to $63.1\%$ when adding modality-level advice (C2), $63.9\%$ when adding explanations instead (C3), and $66.9\%$ when providing both features simultaneously (C4). 

\begin{table}[h!]
  \centering
  \caption{Decomposition of trials into the four outcome categories of \citet{schoeffer2025reliance}. Entries are percentages of all trials (AI correct on $75\%$ of trials).}
  \label{tab:decomposition}
  \small
  \begin{tabular}{@{}lrrrrr@{}}
    \toprule
    & \multicolumn{2}{c}{AI correct ($75\%$)} & \multicolumn{2}{c}{AI incorrect ($25\%$)} & \\
    \cmidrule(r){2-3}\cmidrule(r){4-5}
    Condition & CA & DO & DA & CO & AR \\
    \midrule
    C1 & $50.8$ & $24.2$ & $14.1$ & $10.9$ & $61.8$ \\
    C2 & $54.1$ & $20.9$ & $15.9$ &  $9.1$ & $63.1$ \\
    C3 & $53.9$ & $21.1$ & $15.0$ & $10.0$ & $63.9$ \\
    C4 & $58.0$ & $17.0$ & $16.1$ &  $8.9$ & $66.9$ \\
    \bottomrule
  \end{tabular}

  \vspace{2pt}
  {\footnotesize C1 = Fused, C2 = Decomposed, C3 = Fused Explanation, C4 = Decomposed Explanation. CA = correct adherence, DO = detrimental override, DA = detrimental adherence, CO = corrective override, AR = appropriate reliance.}
\end{table}

To better understand the underlying behavioral mechanisms, we examine how decision-makers interact with the AI advice across trials. Drawing on \citet{schoeffer2025reliance}, we classify each trial into one of four outcomes defined by the intersection of participant action (adherence versus override) and advice correctness (accurate versus inaccurate). Given that the underlying model maintains an accuracy rate of $75\%$ across the stimulus set, these four outcomes partition each condition into an accurate-advice stratum comprising $75\%$ of trials and an inaccurate-advice stratum comprising the remaining $25\%$ (\Cref{tab:decomposition}).

Across the whole dataset, participants override the AI advice in $30.5\%$ of trials, of which $20.8$ percentage points are overrides of correct advice and $9.7$ percentage points are overrides of incorrect advice. Consequently, overriding AI advice is more than twice as likely to introduce human error as it is to correct an AI error.

\begin{figure}[t]
  \centering
  \includegraphics[width=\linewidth]{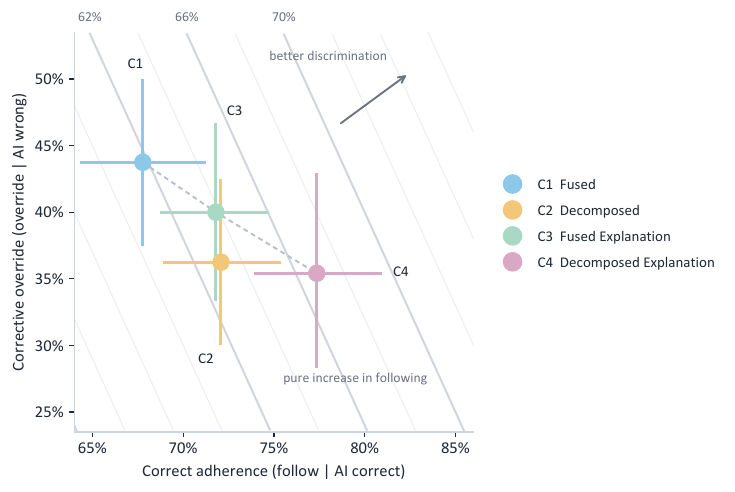}
    \Description{Scatter plot placing the four conditions in a two-dimensional space whose axes are the two components of appropriate reliance. Diagonal gray contours mark constant team accuracy; each condition's team accuracy is printed beside its point, and bars show $95\%$ confidence intervals.}
  \caption{The four conditions in correct-adherence $\times$ corrective-override space. Axes are conditional rates: $x = P(\text{follow}\mid\text{AI correct})$ and $y = P(\text{override}\mid\text{AI incorrect})$. Gray contours are isolines of team accuracy $0.75x + 0.25y$, labeled on the top edge (AI accuracy is 75\% by construction). Rightward movement is a pure increase in following; movement toward the upper right would indicate better discrimination. Bars are 95\% confidence intervals.}
  \label{fig:reliance_space}
\end{figure}

Our conditions differ in how frequently participants adhere to the AI advice, not in how effectively they distinguish correct advice from flawed advice. From C1 to C4, correct adherence increases from $50.8\%$ to $58.0\%$ of all trials, while corrective overriding declines from $10.9\%$ to $8.9\%$, and detrimental adherence expands from $14.1\%$ to $16.1\%$. Expressed as conditional rates within each subgroup, participants adhere to correct advice on $67.8\%$ of AI-correct trials in C1 compared to $77.4\%$ in C4, and they successfully override incorrect advice on $43.8\%$ of AI-incorrect trials in C1 compared to $35.4\%$ in C4. \Cref{fig:reliance_space} plots the four conditions in these two coordinates. The trajectory from C1 to C4 traces a displacement slope of $-0.88$, against $-1.00$ for a uniform increase in adherence to the advice and $-3.00$ for movement along a contour of constant team accuracy, so our conditions differ almost exactly as they would if participants simply agreed with the AI more often across the board. The data show no shift toward the upper right, meaning that richer interfaces do not help participants distinguish correct advice from incorrect advice.

Two pre-registered exploratory tests confirm this pattern. The distribution of outcome categories differs significantly across conditions on AI-correct trials ($\chi^2(3) = 16.65$, $p = .0008$), whereas the AI-incorrect distribution does not differ detectably ($\chi^2(3) = 4.43$, $p = .218$). The experimental manipulations therefore change how often participants correctly adhere to AI advice, without a detectable shift in corrective overriding. Since the AI is accurate on $75\%$ of trials, higher adherence also generates more correct decisions overall, and thus also higher appropriate reliance. The subsections that follow report our pre-registered tests and locate this behavioral pattern within each factor.

\subsection{Model Estimation}
\label{subsec:approach}

We estimate a generalized linear mixed model using a binomial distribution and a logit link function across all $3{,}840$ trial observations. Following our pre-registration, we use advice granularity ($G$), explanation availability ($E$), and modality disagreement ($D$) as dummy-coded fixed factors, along with all two-way and three-way interaction terms. We also add $z$-standardized control covariates for domain expertise, AI familiarity, and propensity to trust, and specify crossed random intercepts for human participants and vehicle listing instances. We report the complete model in \Cref{app:covariates}.

Our model converges with this pre-registered random-effects structure. We estimate participant-level variance at practically zero ($\mathrm{SD} = 0.0004$) alongside substantial instance-level variance ($\mathrm{SD} = 0.929$). Since removing the participant random intercept leaves every fixed-effect estimate identical to $4$ decimal places, we retain the full structure as registered. As an additional check, we aggregate to one accuracy score per participant and rerun the between-subjects comparisons on those $240$ values, where the estimates replicate, with advice granularity $+2.19$ percentage points ($p = .071$), explanations $+2.92$ percentage points ($p = .025$), and $C4$ against $C1$ $+5.10$ percentage points ($p = .011$, $d = 0.42$). We then evaluate our hypotheses through estimated marginal means contrasts from the fitted model, averaging over non-focal factors and applying one-sided tests for the four directional hypotheses. Since H4 hypothesizes that the effect of advice granularity depends on explanation availability, we examine this interaction before interpreting main contrasts pooled across explanation conditions.

\begin{figure}[ht]
  \centering
  \includegraphics[width=\linewidth]{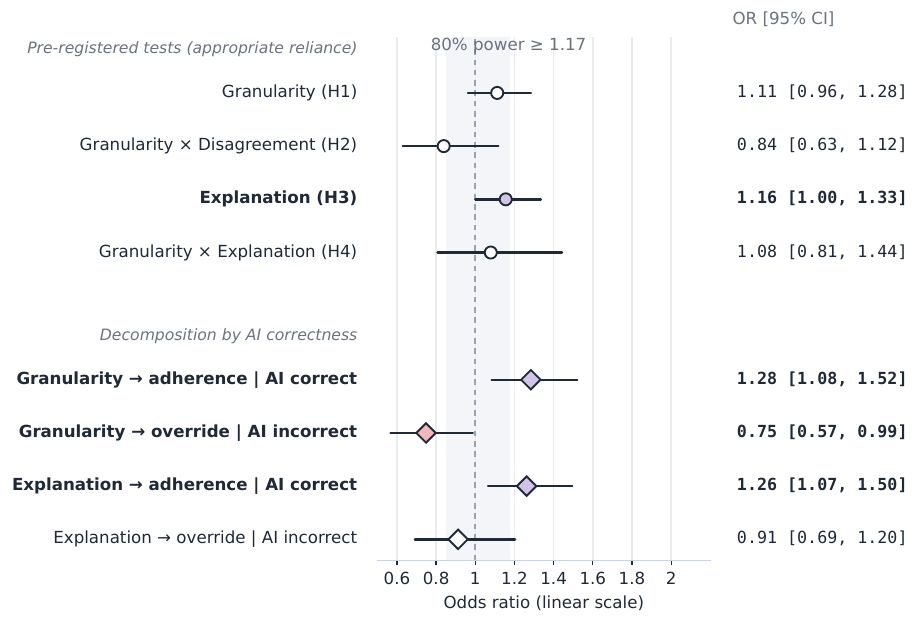}
  \Description{Forest plot of the pre-registered and exploratory estimates on an odds-ratio scale, each with a two-sided $95\%$ confidence interval. A shaded band marks the range of effects our design cannot resolve at $80\%$ power.}
  \caption{Pre-registered and exploratory estimates on an odds-ratio scale. Bars are two-sided $95\%$ confidence intervals; the shaded band marks effects our design cannot resolve at $80\%$ power. H1--H4 are tested one-sided as pre-registered, so H3 is supported ($p = .025$), although its two-sided interval touches one.}
  \label{fig:forest}
\end{figure}

Given our pre-registered sample of $60$ participants per condition and the observed participant-level standard deviation in accuracy of $0.115$ across our $240$ participants, the experiment achieves $80\%$ statistical power to detect a pooled main effect of $3.70$ percentage points and a simple effect of $5.93$ percentage points. Observed differences below these minimal detectable effect sizes fall outside what our sample size can reliably capture, and we treat them as bounds on effect magnitude, not as evidence of absence. \Cref{fig:forest} illustrates each confirmatory and decomposition estimate on a common odds-ratio scale; the shaded band corresponds to the pooled $3.70$ percentage-point bound.

\subsection{Advice Granularity}
\label{subsec:granularity}

With H4, we assume that presenting modality-level advice yields greater performance benefits when accompanied by natural-language explanations. The reasoning is that a rationale attached to each modality lets decision-makers inspect the substance behind each assessment instead of merely noting that the assessments exist. The observed interaction is positive and aligns with the hypothesized direction but falls short of statistical significance ($\beta = 0.076$, $\mathrm{SE} = 0.148$, $z = 0.51$, $p = .304$; $\mathrm{OR} = 1.08$, 95\% CI $[0.81, 1.44]$). Neither simple effect achieves statistical significance on its own, whether explanations are absent ($\beta = 0.068$, $p = .255$) or present ($\beta = 0.144$, $p = .086$). We therefore find no support for H4. Since the interaction effect does not reach significance, we proceed by evaluating the main effect of advice granularity pooled across explanation treatments.

\begin{figure*}[h!]
  \centering
  \includegraphics[width=\linewidth]{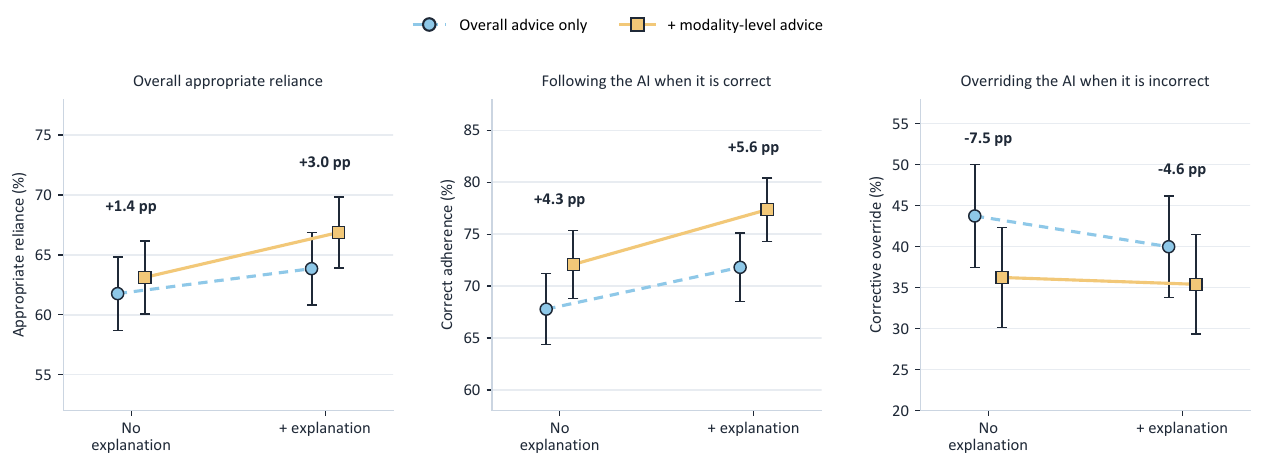}
  \Description{Three-panel figure showing the effect of advice granularity without explanations and with explanations. The left panel plots overall appropriate reliance, the middle panel correct adherence when the AI is right, and the right panel corrective override when the AI is wrong.}
  \caption{Advice granularity without explanations (C1 to C2) and with explanations (C3 to C4).
  Left: overall appropriate reliance. Middle: correct adherence when the AI is right.
  Right: corrective override when the AI is wrong.}
  \label{fig:granularity_pooled}
\end{figure*}

H1 posits that providing modality-level advice increases appropriate reliance. The reasoning is that two assessments resting on different evidence give the decision-maker a cue that varies by case. Where the assessments converge, there are grounds to accept the fused advice, and where they diverge, there are grounds to examine the listing before doing so. Although the point estimate is positive, it does not reach our pre-registered criterion ($\beta = 0.106$, $\mathrm{SE} = 0.073$, $z = 1.44$, $p = .074$; $\mathrm{OR} = 1.11$, 95\% CI $[0.96, 1.28]$). At $2.2$ percentage points pooled, this effect size falls below the $3.70$ percentage points our design is powered to detect. Comparing condition means reveals an accuracy gain of $1.4$ percentage points from C1 to C2 and $3.0$ percentage points from C3 to C4 (\Cref{fig:granularity_pooled,tab:contrasts-appendix}).

To better understand the underlying effects, we decompose this contrast by AI correctness (\Cref{fig:granularity_pooled}). On trials where AI advice is correct, presenting modality-level advice increases adherence from $69.8\%$ to $74.7\%$ ($\beta = 0.250$, $\mathrm{SE} = 0.086$, $z = 2.89$, $p = .004$; $\mathrm{OR} = 1.28$, 95\% CI $[1.08, 1.52]$). Conversely, when AI advice is incorrect, showing modality-level advice reduces corrective overrides from $41.9\%$ to $35.8\%$ ($\beta = -0.290$, $\mathrm{SE} = 0.141$, $z = -2.05$, $p = .040$; $\mathrm{OR} = 0.75$, 95\% CI $[0.57, 0.99]$). Modeling adherence directly is consistent with this reading, as advice granularity does not interact with AI correctness ($\beta = -0.019$, $\mathrm{SE} = 0.165$, $p = .908$). Showing modality-level advice, therefore, makes participants more willing to accept it, regardless of its correctness.

%\textit{Summary:} Revealing modality-level advice made participants more likely to follow the AI, but it did not improve their ability to distinguish correct advice from incorrect advice. The resulting increase in appropriate reliance was therefore driven by greater adherence to correct AI advice, while corrective overriding of incorrect advice declined. We find no evidence that this effect on appropriate reliance depends on whether the modalities agree or disagree.

\subsection{Explanations}
\label{subsec:explanations}

H3 predicts that accompanying advice with natural-language explanations increases appropriate reliance. The reasoning is that a rationale lets decision-makers assess the validity of the underlying logic instead of adhering to or overriding the advice on the basis of how well it matches their prior intuitions. Evaluating this effect across all conditions reveals a positive estimate that reaches statistical significance ($\beta = 0.145$, $\mathrm{SE} = 0.074$, $z = 1.97$, $p = .025$; $\mathrm{OR} = 1.16$, 95\% CI $[1.00, 1.33]$). This means that the odds of a correct decision are about $16\%$ higher when an explanation is present. The gain is demonstrated in both comparisons that vary the availability of explanations, i.e., adding explanations improves appropriate reliance from $C1$ to $C3$ by $2.1$ percentage points, and from $C2$ to $C4$ by $3.8$ percentage points. This second jump is statistically significant on its own ($\beta = 0.183$, $p = .042$). We therefore find support for H3.

\begin{figure*}[h!]
  \centering
  \includegraphics[width=\linewidth]{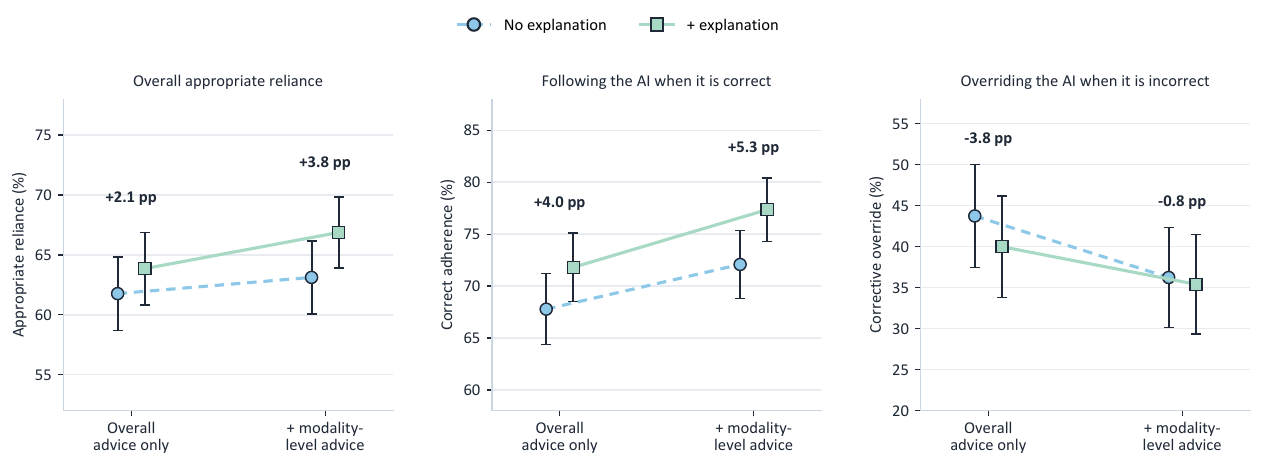}
  \Description{Three-panel figure showing the effect of explanations with fused advice and with decomposed advice. The left panel plots overall appropriate reliance, the middle panel correct adherence when the AI is right, and the right panel corrective override when the AI is wrong.}
  \caption{Explanations with fused advice (C1 to C3) and with decomposed advice (C2 to C4).
  Left: overall appropriate reliance. Middle: correct adherence when the AI is right.
  Right: corrective override when the AI is wrong.}
  \label{fig:explanations_pooled}
\end{figure*}

Next, we decompose appropriate reliance into AI-correct and AI-incorrect trials (\Cref{fig:explanations_pooled}). On trials where AI advice is correct, explanations increase adherence from $69.9\%$ to $74.6\%$ ($\beta = 0.233$, $\mathrm{SE} = 0.087$, $z = 2.69$, $p = .007$; $\mathrm{OR} = 1.26$, 95\% CI $[1.07, 1.50]$). Conversely, when AI advice is incorrect, corrective overriding remains comparable, $40.0\%$ without explanations and $37.7\%$ with explanations ($\beta = -0.092$, $\mathrm{SE} = 0.141$, $z = -0.65$, $p = .515$; $\mathrm{OR} = 0.91$, 95\% CI $[0.69, 1.20]$). The explanation effect on adherence does not differ by AI correctness ($\beta = 0.138$, $\mathrm{SE} = 0.165$, $p = .404$). The performance gain therefore arises without the measurable reduction in corrective overriding that accompanies modality-level advice.

This contrast should not be read as evidence that explanations help participants tell correct advice from incorrect advice. The explanation effect on adherence does not differ between AI-correct and AI-incorrect trials ($\beta = 0.143$, $\mathrm{SE} = 0.165$, $p = .387$). Therefore, the explanation effect on appropriate reliance is not accompanied by a measurable reduction in corrective overriding, whereas the granularity effect is.

%\textit{Summary:} Explanations increased appropriate reliance primarily by increasing adherence when the AI was correct. Unlike modality-level advice, this increase was not accompanied by a measurable reduction in participants' corrective overriding when the AI was wrong. The findings therefore suggest that explanations improved reliance without producing the same broad shift toward AI adherence observed for modality-level advice.

\subsection{Modality Agreement}
\label{subsec:agreement}

H2 predicts that displaying modality-level advice yields greater benefits when the two modalities disagree, as participants who see divergence between the image-based and attribute-based assessments might treat that conflict as a signal and rely less on the fused advice when its inputs point in different directions. Our data show an estimate in the opposite direction, and it fails the pre-registered one-sided test ($\beta = -0.176$, $\mathrm{SE} = 0.147$, $z = -1.20$, $p = .885$; $\mathrm{OR} = 0.84$, 95\% CI $[0.63, 1.12]$). The three-way interaction, for which our pre-registration expected the same conflict-dependence, likewise runs opposite to that expectation ($\beta = -0.317$, $\mathrm{SE} = 0.293$, $z = -1.08$, $p = .280$, two-sided). We therefore find no support for H2: modality-level advice does not help more when the modalities disagree.

\begin{figure*}[h!]
  \centering
  \includegraphics[width=\linewidth]{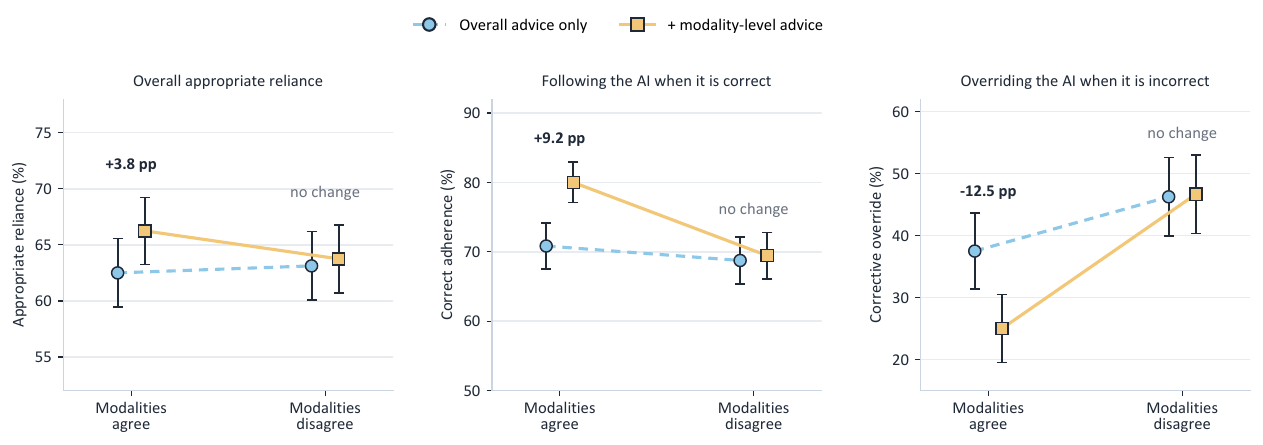}
  \Description{Three-panel figure showing the effect of advice granularity across modality agreement and by AI correctness. The left panel plots overall appropriate reliance, the middle panel correct adherence when the AI is right, and the right panel corrective override when the AI is wrong.}
  \caption{Advice granularity across modality agreement, then by AI correctness.
    Left: overall appropriate reliance. Middle: correct adherence when the AI is right.
    Right: corrective override when the AI is wrong.}
  \label{fig:granularity_pattern}
\end{figure*}

The process measures show where the interface does change behavior (\Cref{fig:granularity_pattern}). On agreement trials, revealing modality-level advice raises appropriate reliance by $3.8$ percentage points; on disagreement trials it does not. Advice granularity and modality disagreement interact on adherence ($\beta = -0.532$, $\mathrm{SE} = 0.148$, $z = -3.59$, $p < .001$), and this interaction does not differ by AI correctness ($\beta = 0.146$, $\mathrm{SE} = 0.332$, $p = .659$). On agreement trials, revealing modality-level advice raises correct adherence from $70.8\%$ to $80.0\%$ and reduces corrective overrides from $37.5\%$ to $25.0\%$. On disagreement trials both remain unchanged ($68.8\%$ versus $69.4\%$ adherence; $46.2\%$ versus $46.7\%$ override). Because AI accuracy is $75\%$ in both strata, agreement carries no information about advice quality, yet it is where the interface effect appears.

A baseline difference between agreement and disagreement trials emerges independently of our manipulation. On incorrect trials, participants override the advice more often when the modalities disagree ($46.5\%$) than when they agree ($31.2\%$). The direction holds in all four conditions, including C1 and C3, where the conflict is hidden behind the fused advice, but the evidence is weak: restricted to those two conditions the difference is $46.2\%$ versus $37.5\%$, and at the stimulus level it rests on eight incorrect listings per stratum ($+15.2$ percentage points, $t = 1.85$, $p = .086$). Car-domain expertise predicts overriding in this subgroup ($\beta = 0.169$, $p = .025$). Our experiment thus supports the narrower claim that visible conflict between modalities does not prompt more corrective overriding, whereas visible agreement suppresses it.

%\textit{Summary:} Visible modality agreement acted as a reassurance cue instead of a diagnostic signal. When the modalities agreed, participants followed the AI substantially more often, including when its advice was wrong; when the modalities disagreed, revealing the breakdown had essentially no effect on adherence or corrective overriding. Thus, the interface made agreement more influential even though agreement contained no information about whether the fused advice was correct.
 
\section{Discussion} \label{sec:discussion}

Following, we discuss our findings and derive implications for research and practice. We then present guidelines for the development of multi-modal AI systems and  derive alternative hypotheses that can serve as avenues for further investigations in future work.

\subsection{Transparency About the System's Reasoning Is Not Evidence About the Case} 
\label{sec:disc-transparency}

An intervention that makes humans adhere more often to an AI system that is correct more often than it is incorrect will increase accuracy, even if it does not help them distinguish correct from incorrect advice. \citet{schoeffer2025reliance} formalize this as a blind-adherence paradox, and our results show the same pattern for a multi-modal system. Revealing modality-level information made our participants more likely to adhere to AI advice, no matter if it was correct or not. Explanations had a similar, yet more targeted effect: they increased adherence when the AI was correct but did not significantly affect overrides when the AI was incorrect. Neither intervention improved participants' ability to distinguish correct from incorrect advice (\Cref{subsec:granularity,subsec:explanations}). This suggests that both interventions increased reliance on the AI's advice without providing participants with complementary ways to assess its accuracy.

One possible explanation for this behavior is that neither form of transparency provided participants with information that helped them determine whether the AI's advice was correct. The verifiability of \citet{fok2024search} similarly suggests that transparency is most useful when humans can use it to confirm or reject an answer. In our study, the modality-level outputs and explanations described how the system reached its advice, but they did not provide additional evidence about the car itself. This is consistent with \citet{poursabzi2021manipulating}, who found that showing how a model arrived at its prediction could make humans less likely to notice large errors because they focused on the model's reasoning, not the case itself. More information about how a system reaches an answer can therefore create a sense of understanding without making the answer itself easier to verify. This also aligns with work showing that placebic explanations can increase trust without adding useful content \citep{eiband2019impact}, and that explanations can lead human to overestimate their perceived understanding of a model \citep{chromik2021think}. Our explanations showed a similar pattern to the feature-importance explanations studied by \citet{schemmer2023appropriate}, where participants increased their reliance on correct advice without reducing their reliance on incorrect advice.

\GuidelineBox{Transparency raises adherence, not error detection}{Revealing modality-level outputs makes humans more likely to follow AI advice. This can improve performance when the AI is more accurate than humans, but it should not be expected to help humans identify incorrect advice. Designers can therefore use agreement to systematically steer reliance, but should do so only when greater adherence is desirable given the relative reliability of the AI and the human.}% If error detection is important, transparency should be paired with case-specific evidence that humans can independently evaluate.}

\subsection{Agreement Is Read as Confirmation of Correctness}
\label{sec:disc-agreement}

The part of the modality-level display that changed behavior was agreement, even though agreement indicated nothing to participants about whether the AI was correct. We held AI accuracy at $75\%$ for both agreeing and conflicting instances, so agreement was uninformative about the correctness of the fused advice, but participants nevertheless used it as a cue to follow the AI. Revealing the outputs increased adherence on agreeing trials and, at the same time, reduced the rate at which participants caught the AI's errors. It had no effect on conflicting trials (\Cref{fig:granularity_pattern}). The hypothesis expected disagreement to be a useful signal, following work in machine learning showing that conflict between modalities can lower a fusion model's confidence in its output \citep{han2021trusted, xu2024reliable}. Instead, participants responded to consensus as if it were evidence.

This response is consistent with the idea that human may treat agreement between sources as support for a conclusion, even when that agreement is not informative. \citet{lu2024second} found the opposite response to visible conflict: a disagreeing second opinion reduced reliance regardless of whether the AI was correct. Although the direction differs from our finding, both results suggest that human may respond to agreement or disagreement without considering whether it actually bears on correctness. Human confidence in an aggregate also tends to increase when its sources agree \citep{budescu2003effects}, and agreement among AI advisors has been shown to increase conformity and overreliance, while a disagreeing advisor can sometimes lead to better decisions \citep{tsuchiya2026more}. In our setting, this creates an important distinction between the perception of confirmation and actual confirmation. The two modalities may have appeared like two sources pointing to the same conclusion, but both came from separate calls to the same model. They could therefore share the same prior knowledge, assumptions, and errors of the modal rather than two different models trained on different data. For example, if both the image and the attribute assessments judged a car to be appropriately priced, this did not provide the same evidential support as two genuinely independent assessments reaching that conclusion. Yet human can treat agreement between dependent sources as independent confirmation, even after being told that only independent support should count as shown by \citep{yousif2019illusion}. This makes a ``herding effect'' one plausible explanation, where humans followed the majority prediction of multi-modal advice and refrained from overriding even when being in doubt. 

Yet, our study cannot establish that a herding caused this effect, and other explanations are possible. For example, agreement may have provided other cues to participants. What the data do show is that agreement increased reliance even though it did not provide information about whether the AI was correct. This is important because most previous work has focused on disagreement, either as a cue that reduces trust or as a way to reduce conformity \citep{lu2024second, tsuchiya2026more}. Our findings suggest that agreement can also shape reliance, particularly in multi-modal systems where different modalities may share the same model and therefore agree without providing independent evidence.

\HypothesisBox{Test why agreement increases reliance}{Future research should establish why modality agreement increases adherence and whether that response can be redirected. Manipulating source independence, the number of agreeing outputs, and whether agreement predicts correctness would separate herding from perceived corroboration and show how far the response is open to influence.}

\subsection{Conflict Is Visible and Goes Unused}
\label{sec:disc-conflict}

While agreement increased adherence, disagreement did not change participants' decisions. Revealing modality-level outputs had no effect on disagreement trials. At the same time, participants tended to be more likely to override incorrect AI advice when modalities disagreed than when they agreed. This pattern suggests that disagreement contained information relevant to identifying AI errors, yet participants did not make greater use of that information when it was revealed. Because the direction appears in C1 and C3 as well, the listings themselves may have made these cases easier to recognize as problematic, but this baseline difference means that the effect was not coming from revealing modality-level advice. This is consistent with evidence that uni-modal AI and multi-modal AI models tend to diverge on cases where human annotators themselves are uncertain \citep{srikanth2025mixed}. The disagreement between modalities, therefore, contained a potentially useful signal, but the interface did not help participants use it.

The absence of an effect of displayed disagreement may have several explanations. One possibility is that the interface showed disagreement without explaining what it meant. Participants could see that the modalities disagreed, but were given no reason to treat this as a warning. Another possibility is that the disagreement itself may not have appeared equally meaningful across the two modalities. In our study, the image-based assessment tended to favor an appropriate price, so a disagreement with the attribute-based assessment may not have been perceived as a strong or informative conflict. The explanations did not resolve this either, because each one described only its own modality. The image explanation focused on visible condition, while the attribute explanation focused on the vehicle specifications. \citet{miller2023explainable} argues that useful explanations should help explain why one outcome occurred instead of another.  \citet{reingold2024dissenting} further found that adding per-model attributions to a simple disagreement signal could increase overreliance. These possibilities make it unclear whether participants ignored disagreement because they did not know how to interpret it, because the particular pattern of disagreement was not informative enough, or both.

\HypothesisBox{Test when is conflict informative}{Future research should understand how humans perceive and interpret modality conflict before testing how to make it actionable.}

\subsection{Steering Overriding Needs a Reliability Signal}
\label{sec:disc-gating}

Our results suggest that transparency's effect on reliance depends on how reliable the AI is relative to the human using it. In our study, both transparency interventions increased adherence to an AI that was more likely to be correct than the average participant. This increased accuracy since participants adhered to the more reliable source more often, but it did not help them distinguish correct from incorrect advice (\Cref{subsec:granularity,subsec:explanations}). If humans were more reliable than the AI, the same increase in adherence could instead reduce accuracy. Transparency that simply shifts reliance toward the AI is therefore not necessarily beneficial; its value depends on whether the resulting shift moves reliance toward the more reliable source.

This makes information about the AI's reliability particularly important. Prior work has shown benefits from communicating estimated correctness\citep{ma2023who}, confidence\citep{kim2024notsure}, or uncertainty \citep{marusich2024using}. Such information can help humans judge when the AI's advice deserves more or less weight \citep{Spitzerdelegation2025}. In a multi-modal system, disagreement between modalities could provide another such signal if it is associated with lower reliability. For example, modality-level outputs could be restricted to be only shown  when the modalities disagree to avoid the effect of increased adherence on agreeing modalities while retaining disagreement as a potential warning signal. Yet,  this requires disagreement of modalities to be sufficiently predictive of errors to serve this purpose remains an empirical question as well as additional information that actually drives overriding of AI advice.

In our study, modality disagreement was presented as a fact about the system, but participants were left to decide whether it was important. Interventions that have improved discrimination have instead communicated signals such as correctness likelihood \citep{ma2023who}, the AI system's accuracy patterns \citep{Spitzerdelegation2025} or own uncertainty \citep{kim2024notsure}, or instance-level uncertainty \citep{marusich2024using}. If disagreement is shown to predict lower reliability, communicating this relationship more directly may help humans make better use of the signal. A stronger version could also explain why the modalities disagree, such as why the image- and attribute-based assessments point in different directions instead of simply explaining each modality separately.

These possibilities all aim to address the same limitation: increased adherence is useful only when the AI is the more reliable source.. The broader design question is therefore how to connect transparency to signals of reliability, so that additional information supports scrutiny when it is needed instead of simply increasing reliance.

\HypothesisBox{Identify ways to Steer Overriding through Modality Disagreement}{Future research should test whether pairing modality conflict with a reliability signal such as explaining the disagreement or providing the system's confidence prompts overriding AI advice mroe often.}

\subsection{Limitations and Future Research}
\label{sec:disc-limitations}

Since we held AI accuracy at 75\% in both the agreement and disagreement subgroups, we could show that participants treated agreement as reassuring even when it did not indicate whether the AI was correct. This design cannot tell us whether they would have used disagreement as a warning had it predicted errors, which it can in deployed systems \citep{han2021trusted, srikanth2025mixed}. Participants also received no outcome feedback, so they had no opportunity to learn whether disagreement was a useful cue, even though reliance changes as humans observe system accuracy \citep{yin2019understanding}.

Finally, the binary task limits what we can learn about how participants used the AI. A continuous price judgment, for example, could reveal changes in the size of participants' estimates or their confidence that a binary decision cannot capture. Our participants were non-professional used-car buyers, and domain expertise can change how humans respond to AI advice \citep{logg2019algorithm, gaube2021susceptibility}. The findings may likewise differ in other markets, domains, or populations. Finally, we did not record whether participants viewed the modality-level outputs on each trial, so we cannot distinguish ignoring a visible cue from not attending to it.

\section{Conclusion}
\label{sec:disc-conclusion}

In this work, we investigate whether decomposing multi-modal AI advice into modality-level outputs helps humans rely on AI advice more appropriately. Exposing the system's internal structure increased overall adherence without increasing error detection. Both modality-level advice and natural-language explanations led participants to follow the AI more frequently, regardless of whether its advice was correct. Consequently, accuracy gains in our richest condition reflected higher baseline reliance on the model but did not reflect the ability to discriminate between sound and flawed advice. 

Further analysis of modality agreement explains this shift in reliance: visible agreement between modalities acted as a persuasive cue, significantly boosting adherence even though agreement provided no signal of the actual correctness of the advice. Conversely, visible disagreement did not prompt corrective overrides. These findings demonstrate that displaying internal model assessments can make AI advice more persuasive without making its failure modes easier to detect.

Our study is one of the first explorations of how humans respond to modality-level AI advice, and the central design question remains open: how to connect what a system reveals about itself to the reliability of a specific piece of advice. Whether modality disagreement can serve that role depends on whether it tracks AI failure and whether humans register it as a warning when it does. Future research should therefore explore these open questions and identify ways to make modality-level advice help humans catch errors instead of only making them trust the AI more.

\bibliographystyle{ACM-Reference-Format}
\bibliography{sample-base}

\clearpage
%TC:ignore

% =============================================================================
%   app:dataset_sampling, app:ai_pipeline, tab:prompt_templates,
%   app:sampling_tables, tab:stimulus_pool_distribution,
%   tab:participant_trial_sampling, tab:all_measures, app:covariates,
%   tab:contrasts, tab:contrasts-appendix
% =============================================================================

\section*{Appendix}
\label{sec:appendix}

\subsection{Participant Flow and Exclusions}
\label{app:exclusions}

We recruited adults via Prolific who resided in Germany or Austria, were fluent in German, and reported prior experience purchasing or evaluating used vehicles. Professional car dealers and automotive appraisers were ineligible. After consent and a guided tutorial, participants completed a comprehension check (up to two attempts). Those who failed twice were returned to Prolific without completing the main task. Two instructed-response attention items were embedded in the pre- and post-study questionnaires; a hidden honeypot field in the post-study form flagged automated submissions. We additionally excluded incomplete main-task sessions.

\begin{table}[h]
  \centering
  \small
  \caption{Participant flow into the analysis sample.}
  \label{tab:participant_flow}
  \begin{tabular}{@{}lr@{}}
    \toprule
    Stage & $n$ \\
    \midrule
    Failed comprehension check twice (returned) & 12 \\
    Incomplete main task ($<$16 trials) & 3 \\
    Failed both instructed-response items & 0 \\
    Flagged by post-study honeypot & 0 \\
    \midrule
    Final analysis sample & 240 \\
    \quad per between-subjects condition (C1--C4) & 60 \\
    Trial observations ($240 \times 16$) & 3{,}840 \\
    \bottomrule
  \end{tabular}
\end{table}

The final sample ($N = 240$) had mean age $35.4$ years ($SD = 11.7$, range $18$--$75$); $65.9\%$ male, $33.6\%$ female, $0.5\%$ non-binary / prefer not to say. Most resided in Germany ($91.4\%$), the remainder in Austria; German was the first language of $69.4\%$. Median completion time was $17.0$ minutes (fixed payment \pounds2.50; effective hourly rate \pounds8.82). One participant failed a single instructed-response item and was retained as pre-registered.

\subsection{Dataset Sampling Criteria}
\label{app:dataset_sampling}

To ensure high data quality, listings scraped from mobile.de had to meet strict eligibility thresholds before entering the candidate pool. We required listings to include at least six high-resolution photographs and at least four of five core specifications, namely registration year, mileage, fuel type, vehicle brand, and model. We discarded listings that lacked critical price indicators or contained corrupted image files prior to expert evaluation.

\subsection{AI Generation Pipeline and Validation Protocols}
\label{app:ai_pipeline}

We deployed GPT-5.1 via the official API with temperature set to~$0$ to maximize deterministic consistency across runs. System prompts instructed the model to evaluate vehicle listings from the perspective of an expert certified automotive appraiser operating in the German used-car market. Within the generation pipeline, prompts classified listings using the binary categories \emph{Reasonable} and \emph{Overpriced}, matching the labels shown on the participant interface.

For each listing, we issued three independent API calls so that modality-level evaluations were not post-hoc algorithmic splits of a single fused run: (i)~image-only (asking price + up to four evenly sampled photographs); (ii)~attribute-only (asking price + structured specifications, withholding platform ratings and dealer identities); (iii)~combined (photographs, specifications, and asking price). Each call returned a binary advice label and a concise natural-language explanation. Image explanations were required to draw only on visible vehicle condition; attribute explanations only on technical specifications; combined explanations were forbidden from contrast/conflict language that would reveal disagreement.

To maintain strict boundaries between modalities, we enforced automated filtering scripts on all generated explanations before accepting them into the study pool. These scripts restricted explanation length to between 10 and 60 words across a maximum of two sentences. For image-only queries, a forbidden-phrase filter prohibited terms related to structured data fields such as exact mileage figures, registration years, or fuel classifications. Conversely, attribute-only queries triggered filters prohibiting visual terminology such as paint condition, cosmetic wear, denting, or interior cleanliness. If a generated explanation violated length constraints or triggered a forbidden-phrase filter, our pipeline initiated automated regeneration attempts up to a maximum of three iterations before flagging the listing for manual inspection. The complete prompt templates for all three query streams appear in \Cref{tab:prompt_templates}.

\begin{table}[t]
\centering
\small
\caption{Prompt templates for the three AI query streams
  (system role + human instruction; temperature $=0$).
  Pipeline labels use \textsc{Reasonable}/\textsc{Overpriced}, matching the participant UI.}
\label{tab:prompt_templates}
\setlength{\tabcolsep}{3pt}
\begin{tabular}{@{}p{1.4cm}p{2.9cm}p{3.7cm}@{}}
\toprule
Stream & System role (summary) & human instruction (summary) \\
\midrule
Image-only
  & German used-car appraiser; photos + asking price only; no written specs.
  & Decide Reasonable/Overpriced from visual condition vs.\ asking price.
    Output \texttt{advice} + one-sentence \texttt{EXPLANATION}.
    Forbidden: photo-meta phrasing; attribute terms (mileage, year, fuel, \ldots). \\
Attribute-only
  & Same role; written specs + price only; no photographs.
  & Decide Reasonable/Overpriced from attributes vs.\ market value.
    Output \texttt{advice} + one-sentence \texttt{EXPLANATION}.
    Forbidden: visual language (looks/appears, scratch, dent, paint, \ldots). \\
Combined
  & Same role; photographs, specifications, and asking price.
  & Decide Reasonable/Overpriced from photos + attributes.
    Output \texttt{advice} + one--two sentence \texttt{EXPLANATION}.
    Forbidden: contrast/conflict language and source-comparison phrasing
    (however, despite, ``the photo'', ``the attributes'', \ldots). \\
\bottomrule
\end{tabular}
\end{table}

\begin{table}[h]
\centering
\small
\setlength{\tabcolsep}{3pt}
\caption{Distribution of the 64-instance stimulus pool across modality agreement, ground truth, and AI correctness.}
\label{tab:stimulus_pool_distribution}
\resizebox{\linewidth}{!}{%
\begin{tabular}{@{}|l|l|cc|c|c|@{}}
\hline
\textbf{Modality} & \textbf{Ground Truth} & \textbf{Correct} & \textbf{Incorrect} & \textbf{Total} & \textbf{Accuracy} \\
\hline
\multirow{2}{*}{\textbf{Agreement}}
  & Reasonable  & 12 & 4 & 16 & $75\%$ \\
  & Overpriced  & 12 & 4 & 16 & $75\%$ \\
\hline
\multirow{2}{*}{\textbf{Disagreement}}
  & Reasonable  & 12 & 4 & 16 & $75\%$ \\
  & Overpriced  & 12 & 4 & 16 & $75\%$ \\
\hline
\textbf{Total Pool} & & \textbf{48} & \textbf{16} & \textbf{64} & \textbf{$75\%$} \\
\hline
\end{tabular}%
}
\end{table}

\begin{table}[h]
\centering
\small
\setlength{\tabcolsep}{3pt}
\caption{Sampling matrix of the 16 decision trials completed by each participant.}
\label{tab:participant_trial_sampling}
\resizebox{\linewidth}{!}{%
\begin{tabular}{@{}|l|l|cc|c|c|@{}}
\hline
\textbf{Modality} & \textbf{Ground Truth} & \textbf{Correct} & \textbf{Incorrect} & \textbf{Total} & \textbf{Accuracy} \\
\hline
\multirow{2}{*}{\textbf{Agreement}}
  & Reasonable  & 3 & 1 & 4 & $75\%$ \\
  & Overpriced  & 3 & 1 & 4 & $75\%$ \\
\hline
\multirow{2}{*}{\textbf{Disagreement}}
  & Reasonable  & 3 & 1 & 4 & $75\%$ \\
  & Overpriced  & 3 & 1 & 4 & $75\%$ \\
\hline
\textbf{Per Participant} & & \textbf{12} & \textbf{4} & \textbf{16} & \textbf{$75\%$} \\
\hline
\end{tabular}%
}
\end{table}

\subsection{Measures and Formalizations}
\label{app:measures}

\Cref{tab:all_measures} summarizes all behavioral and questionnaire measures. English wording is shown for readability; participants completed German translations of the same items.

\begin{table*}[t]
  \centering
  \small
  \caption{Behavioral and questionnaire measures used in the experiment.}
  \label{tab:all_measures}
  \begin{tabular}{@{}p{0.17\textwidth}p{0.48\textwidth}p{0.27\textwidth}@{}}
    \toprule
    Measure & Operationalization & Scale / coding \\
    \midrule
    Appropriate reliance ($AR$)
      & Final decision matches expert ground truth.
        In this binary task with fixed AI advice, $AR$ is equivalent to
        human--AI team performance~\citep{schemmer2022should, schoeffer2025reliance}.
      & Binary trial indicator \\[4pt]
    Correct adherence (CA)
      & Participant follows AI when AI is correct
      & Binary; defined on AI-correct trials \\
    Corrective override (CO)
      & Participant overrides AI when AI is incorrect
      & Binary; defined on AI-incorrect trials \\
    Detrimental adherence (DA)
      & Participant follows AI when AI is incorrect
      & Binary; defined on AI-incorrect trials \\
    Detrimental override (DO)
      & Participant overrides AI when AI is correct
      & Binary; defined on AI-correct trials \\[4pt]
    Adherence
      & Participant's verdict equals the AI's overall advice
      & Binary trial indicator \\
    Trial confidence
      & Self-reported confidence after each decision
      & 5 verbal levels $\mapsto$ $\{1,\ldots,5\}$ \\
    Time on page
      & Seconds spent on the listing screen (minimum enforced 20\,s)
      & Continuous (seconds) \\
    \midrule
    Car-domain expertise
      & ``How familiar are you with assessing the prices of used cars?''
      & 1--5 Likert; $z$-scored covariate \\
    AI familiarity
      & Mean of three items (facets; decision knowledge; proficiency)
      & 1--5 Likert mean; $z$-scored \\
    Propensity to trust AI
      & Mean of five positively keyed items and one reverse-coded distrust item
        (Merritt-style propensity~\citep{merritt2013trust})
      & 1--5 composite; $z$-scored \\
    Cognitive load
      & Mean of NASA-TLX--style items: mental demand, effort, frustration~\citep{hart1988development}
      & 1--5 Likert mean (post-task) \\
    Post-task confidence
      & ``How confident were you in your judgments overall?''
      & 1--5 Likert \\
    AI influence
      & ``How much did the AI influence your decisions?''
      & 1--5 Likert \\
    \bottomrule
  \end{tabular}
\end{table*}

\subsection{Statistical Model and Covariates}
\label{app:covariates}

\textit{Primary model.}
We estimate a binomial generalized linear mixed model with logit link on all $N_{\text{trials}}=3{,}840$ observations (R packages \texttt{lme4} and \texttt{emmeans}):
\begin{align*}
  \mathrm{logit}\bigl(\Pr(AR_{ij}=1)\bigr)
  &= \beta_0
    + \beta_G G_i
    + \beta_E E_i
    + \beta_D D_{ij} \\
  &\quad
    + \beta_{GE}(G_i E_i)
    + \beta_{GD}(G_i D_{ij}) \\
  &\quad
    + \beta_{ED}(E_i D_{ij})
    + \beta_{GED}(G_i E_i D_{ij}) \\
  &\quad
    + \gamma_1\,\mathrm{Expertise}_i^{z}
    + \gamma_2\,\mathrm{AIFam}_i^{z}
    + \gamma_3\,\mathrm{Trust}_i^{z} \\
  &\quad
    + u_{j}
    + v_{i},
\end{align*}
where $i$ indexes participants and $j$ listings; $G{=}1$ for modality-level advice (C2, C4); $E{=}1$ for explanations (C3, C4); $D{=}1$ for modality disagreement; $u_j\sim\mathcal{N}(0,\sigma_u^2)$ is a listing random intercept ($\mathrm{SD}=0.929$); and $v_i\sim\mathcal{N}(0,\sigma_v^2)$ is a participant random intercept. Participant-level variance was essentially zero ($\mathrm{SD}\approx 0.0004$). Removing $v_i$ left every fixed-effect estimate unchanged to four decimal places, so we retain the crossed structure as pre-registered. Hypotheses H1--H4 are evaluated as one-sided estimated-marginal-mean contrasts ($>$); the three-way interaction is reported as a two-sided exploratory test.

\textit{Sensitivity.}
With $60$ participants per condition and observed participant-level $SD$ of accuracy $0.115$, the design achieves approximately $80\%$ power for a pooled main effect of $3.70$ percentage points and an interaction or simple effect of $5.93$ percentage points. Effects below these bounds are treated as magnitude bounds, not evidence of absence.

\textit{Preregistration.}
Hypotheses, design, and analysis plan were pre-registered on AsPredicted (\url{https://aspredicted.org/ds5b2y.pdf}). There is no deviation from the registered random-effects structure: we retain crossed participant and listing intercepts. The participant intercept is estimated at essentially zero variance, and omitting it does not change the fixed-effect estimates.

\subsection{Confirmatory Contrasts and Full Coefficient Table}
\label{app:model_tables}

\needspace{6cm}
\begin{minipage}{\linewidth}
\begin{center}
\small
\captionof{table}{Pre-registered hypothesis tests from the generalized linear mixed model
    ($N = 240$, 3{,}840 trials).
    $p$-values are one-sided for H1--H4 following the pre-registration;
    the three-way term is two-sided exploratory.
    $\beta$ are log-odds and $\mathrm{OR} = e^{\beta}$.}
\label{tab:contrasts}
\setlength{\tabcolsep}{2pt}
\begin{tabular}{@{}llrrrrl@{}}
    \toprule
    & Contrast & $\beta$ & SE & $z$ & $p$ & OR [95\% CI] \\
    \midrule
    H1 & $G$                   &  0.106   & 0.073 &  1.44   & $.074$     & 1.11 [0.96, 1.28] \\
    H2 & $G \times D$          & $-0.176$ & 0.147 & $-1.20$ & $.885$     & 0.84 [0.63, 1.12] \\
    H3 & $E$                   &  0.145   & 0.074 &  1.97   & $.025^{*}$ & 1.16 [1.00, 1.33] \\
    H4 & $G \times E$          &  0.076   & 0.148 &  0.51   & $.304$     & 1.08 [0.81, 1.44] \\
    (expl.) & $G \times E \times D$ & $-0.317$ & 0.293 & $-1.08$ & $.280$ & 0.73 [0.41, 1.29] \\
    \bottomrule
\end{tabular}
\end{center}
\end{minipage}
\par\medskip
\needspace{5cm}
\begin{minipage}{\linewidth}
\begin{center}
\small
\captionof{table}{Condition-level contrasts from the model reported in \Cref{tab:contrasts}.
    One-sided $p$-values on the log-odds scale. ${}^{*}p < .05$.}
\label{tab:contrasts-appendix}
\setlength{\tabcolsep}{2pt}
\begin{tabular}{@{}lrrrr@{}}
    \toprule
    Contrast & $\beta$ & SE & $z$ & $p$ \\
    \midrule
    C2 $-$ C1 ($G$ alone)     & 0.068 & 0.103 & 0.66 & $.255$ \\
    C4 $-$ C3 ($G$ given $E$) & 0.144 & 0.105 & 1.37 & $.086$ \\
    C3 $-$ C1 ($E$ alone)     & 0.107 & 0.103 & 1.04 & $.150$ \\
    C4 $-$ C2 ($E$ given $G$) & 0.183 & 0.106 & 1.73 & $.042^{*}$ \\
    C4 $-$ C1 (combined)      & 0.247 & 0.104 & 2.37 & $.009^{*}$ \\
    Linear trend (C1$\to$C4)  & 0.196 & 0.082 & 2.38 & $.009^{*}$ \\
    \bottomrule
\end{tabular}
\end{center}
\end{minipage}
\par\medskip
\needspace{9cm}
\begin{minipage}{\linewidth}
\begin{center}
\footnotesize
\captionof{table}{Full mixed-effects logistic model for appropriate reliance
    ($N = 240$ participants, $3{,}840$ trials, $64$ listings).
    Binomial GLMM with logit link (\texttt{lme4::glmer}).
    Dummy coding: $G{=}1$ for modality-level advice (C2, C4); $E{=}1$ for explanations (C3, C4); $D{=}1$ for modality disagreement.
    Do \emph{not} read H1/H3 from the raw $G$/$E$ rows; use \Cref{tab:contrasts}.}
\label{tab:glmm-coefs}
\setlength{\tabcolsep}{2pt}
\begin{tabular}{@{}lrrrrc@{}}
    \toprule
    Term & $\beta$ & SE & $z$ & $p$ & OR [95\% CI] \\
    \midrule
    Intercept & $0.597$ & $0.195$ & $3.06$ & $.002^{*}$ & $1.82$ $[1.24, 2.66]$ \\
    $G$ & $0.077$ & $0.149$ & $0.52$ & $.606$ & $1.08$ $[0.81, 1.45]$ \\
    $E$ & $0.005$ & $0.148$ & $0.03$ & $.974$ & $1.00$ $[0.75, 1.34]$ \\
    $D$ & $-0.094$ & $0.274$ & $-0.34$ & $.731$ & $0.91$ $[0.53, 1.56]$ \\
    Car expertise ($z$) & $0.012$ & $0.039$ & $0.30$ & $.766$ & $1.01$ $[0.94, 1.09]$ \\
    AI familiarity ($z$) & $0.051$ & $0.039$ & $1.31$ & $.192$ & $1.05$ $[0.97, 1.14]$ \\
    Trust propensity ($z$) & $0.053$ & $0.038$ & $1.41$ & $.160$ & $1.05$ $[0.98, 1.14]$ \\
    $G \times E$ & $0.234$ & $0.213$ & $1.10$ & $.272$ & $1.26$ $[0.83, 1.92]$ \\
    $G \times D$ & $-0.018$ & $0.205$ & $-0.09$ & $.932$ & $0.98$ $[0.66, 1.47]$ \\
    $E \times D$ & $0.203$ & $0.206$ & $0.99$ & $.323$ & $1.23$ $[0.82, 1.83]$ \\
    $G \times E \times D$ & $-0.317$ & $0.293$ & $-1.08$ & $.280$ & $0.73$ $[0.41, 1.29]$ \\
    \midrule
    \multicolumn{6}{@{}l}{\textit{Random effects}} \\
    Listing SD & $0.929$ & & & & Var. $= 0.863$ \\
    Participant SD & $\approx 0.0004$ & & & & Retained (singular) \\
    \bottomrule
\end{tabular}

\vspace{4pt}
{\footnotesize Fixed-effect $p$-values are two-sided Wald tests. $\beta$ are log-odds; $\mathrm{OR} = e^{\beta}$; 95\% CIs are Wald intervals on the odds-ratio scale. Baseline (intercept) is C1 on agreement trials at mean covariate values.}
\end{center}
\end{minipage}

\subsection{Language and Materials}
\label{app:language}

Because listings came from the German market, the study was conducted in German: listing attributes, AI advice, explanations, instructions, tutorial, comprehension checks, and questionnaires were all presented in German. The English item wordings in this appendix are the source translations used for documentation. Interface labels referred to photograph-based and specification-based AI evaluations (corresponding to the image and attribute streams in the analysis).

%TC:endignore 
\clearpage 
\end{document}